\documentclass[]{article}  

\usepackage{graphics}

\usepackage{amsmath}

\usepackage{makeidx}     
\usepackage{graphicx}    
 \usepackage{amssymb}                        
\usepackage{multicol}    
\usepackage{epsfig}

\usepackage{amssymb}

\begin{document}

\title{\Large{\textsc{\bf High-Gain Nonlinear Observer for Simple Genetic Regulation Process}} }  

\author{L.A. Torres$^1$\footnote{E-mail: ltorres@ipicyt.edu.mx}, V. Ibarra-Junquera$^2$\footnote{E-mail: vij@ucol.mx},
P. Escalante-Minakata$^1$\footnote{E-mail: minakata@ipicyt.edu.mx},
and  H.C. Rosu$^1$\footnote{E-mail: hcr@ipicyt.edu.mx; corresponding
author.$\qquad$ calis07b.tex} 
}


\date{
 $^1$ \small  Instituto Potosino de Investigaci\'on Cient\'{\i}fica y Tecnol\'ogica\\ A.P. 3-74 Tangamanga,
San Luis Potos\'{\i}, SLP 78216, M\'exico\\
$^2$ \small Faculty of Chemical Sciences, University of Colima\\ Coquimatl\'an, Col 28400, Mexico\\
}    %
\maketitle

\begin{center}
published: Physica A 380 (1 July 2007) 235-240\\
doi: 10.1016/j.physa.2007.02.105\\
arXiv:q-bio/0409036v2 [q-bio.QM]
\end{center}

\bigskip

\begin{abstract}
\noindent High-gain nonlinear observers occur in the nonlinear
automatic control theory and are in standard usage in chemical
engineering processes. We apply such a type of analysis in the
context of a very simple one-gene regulation circuit. In general, an
observer combines an analytical differential-equation-based model
with partial measurement of the system in order to estimate the
non-measured state variables. We use one of the simplest observers,
that of Gauthier et al., which is a copy of the original
system plus a correction term which is easy to calculate. For the
illustration of this procedure, we employ a biological model,
recently adapted from Goodwin's old book by De Jong, in which one
plays with the dynamics of the concentrations of the messenger RNA
coding for a given protein, the protein itself, and a single
metabolite. Using the observer instead of the metabolite, it is
possible to rebuild the non-measured concentrations of the mRNA  and
the protein.

\medskip

\noindent {\em  Keywords:} High-gain observer; Diffeomorphism; Gene
regulation; Gene expression

\end{abstract}

\newpage

\section{Introduction}
According to textbooks, gene expression is a very complicated
dynamical process which is regulated at a number of its stages
during the synthesis of proteins \cite{Lewin99}. Similar to many big
cities, with heavy traffic, biological cells host complicated
traffic of biochemical signals at all levels.
At the nanometer scale, clusters of molecules in the form of
proteins drive the dynamics of the cellular network that
schematically can be divided into four regulated parts: the DNA or
genes, the transcribed RNAs, the set of interacting proteins, and
the metabolites. Genes can only affect other genes through specific
proteins, as well as through some metabolic pathways that are
regulated by proteins themselves. They act to catalyze the
information stored in DNA, all the way from the fundamental
processes of transcription and translation to the final
quantities of produced proteins.\\

For the purpose of modeling, it is essential to generate simple
models that help to understand elementary dynamical components of
these complex regulatory networks as molecular tools that
participate in an important way in the machinery of cellular
decisions, that is to say, in the behavior and genetic program of
cells. The central importance of control theory in Biology can be
assessed through the recent problem of identifying control motifs
(or modules), which are patterns that occur in a gene network far
more often than in randomized networks of biological regulators
\cite{mangan03}. This hot issue has been first pinpointed in a
breakthrough paper of Doyle and collaborators \cite{jdoyle} in which
the regulation of bacterial chemotaxis was interpreted in terms of
the simple integral control `adaptive module' introduced by Barkai
and Leibler \cite{BL97}. Since gene regulation appears to occur only
at some definite states of the whole process, which in general are
not well known, we are from the point of view of control engineering
in the case of the reconstruction of those specific states under the
condition of
limited information.\\

It is quite clear that the availability of all state variables to
direct measurement is an extremely rare occasion for gene expression
phenomena or when it is possible it could be too expensive. For this
particular task, but in completely different technological areas,
the engineers have developed software sensors (state observers) that
accurately reconstruct the state variables of various technological
processes \cite{Stephanopoulos}. The basic concept of state of a
system or process could have many different empirical meanings in
biology. For the particular case of gene expression, the meaning of
a state is essentially that of a concentration. The typical problem
in control engineering that appears to be tremendously useful in
biology is the reconstruction of some specific regulated states
under conditions of limited information. \\

In general, an observer is expected to provide a good estimate
$\hat{X}(t)$ of the natural state $X(t)$ of the original system. For
this, one usually can think that some distance
$d\left(\hat{X}(t),X(t)\right)$ (in the sense of a norm $\|\cdot \|$
in a vectorial space) goes to zero as $t \rightarrow \infty$. Such
softwares can be constructed using the mathematical model of the
process to obtain an estimate $\hat{X}$ of the true state $X$. This
estimate can then be used as a substitute for the unknown state $X$.
The usage of state observers has proven useful in process monitoring
and for many other tasks. The concept of observer is used herein in
the sense of control theory, defining an algorithm capable of giving
a reasonable estimation of the unmeasured variables of a process. In
the case of gene expression processes the description is made very
concrete in the following by looking at quite simple mathematical
models that refer to single gene cases and which in principle can be
extended to
some \emph{operons} that are single gene clusters.\\

In this paper we will examine in detail a particularly simple
observer due to Gauthier and collaborators \cite{J.P. Gauthier 92}
possessing arbitrary exponential decay and linear error dynamics for
the case of a three-state genetic regulation process. We were led to
consider this observer because of its simplicity and its {\em high
gain} property. The gain is defined as the amount of increase in
error in the dynamics of the observer. This amount is directly
related to the velocity with which the observer recovers the unknown
signal. For the observer of Gauthier {\em et al} the amount of
increase in error is constant and usually of high values leading to
a fast recover of the unmeasurable states.

\section{Mathematical Model for a Simple Gene Regulation Process}
A kinetic model of a simple genetic regulation process was first
developed by Goodwin as long ago as 1963 \cite{Goodwin63}. It has
been further generalized by Tyson and Othmer \cite{Tyson78} and
clearly explained by De Jong in his recent review \cite{DeJong02}.
We consider here the most simple version of this kinetic model. For
three concentrations $X_1$, $X_2$, $X_3$, corresponding to the
messenger RNA (mRNA) that codes for the unstable enzyme, the enzyme,
and the metabolite, respectively, we write Tyson's model in the form

\begin{eqnarray}
\Gamma _{\rm biology}: \ \ \ \left \{
\begin{array}{c}
\dot{X}_1 = K_{1}\ H\left(X,\vartheta\right)-{\gamma}_{1}{X}_{1} \label{Eqn1}\\
\dot{X}_2 = K_{2}X_1-{\gamma}_{2}{X}_{2}\label{Eqn2}\\
\dot{X}_3 = K_{3}X_2-{\gamma}_{3}{X}_{3}~.\label{Eqn3}
\end{array}\right.
\end{eqnarray}
The parameters $K _1, K _2, K _3$ are all strictly positive and
represent production constants, whereas $\gamma _1, \gamma _2,
\gamma _3$ are also strictly positive degradation constants. These
rate equations express a balance between the number of molecules
appearing and disappearing  per unit time. Notice that the model
assumes that the concentration $X_2$ increases linearly with $X_1$
and the concentration $X_3$ linearly with $X_2$, which are natural
assumptions. In the case of $X_1$, the first term is the production
term involving a nonlinear nondissipative {\em regulation function}
$H$ that we take of the $m$-steepen Hill form ($m>0$ is the
steepness parameter) in common use

\begin{eqnarray}
H^{+}\left(X,\vartheta\right) &=& \frac{X^{m}_3}{X^{m}_3+\vartheta^m}~, \nonumber\\
\\
H^{-}\left(X,\vartheta\right) &=& 1-
\frac{X^{m}_3}{X^{m}_3+\vartheta^m}~, \nonumber
\end{eqnarray}
for the activation and inhibition cases, respectively. The parameter
$\theta$ gives the threshold for the regulatory influence of the
concentration of the metabolite on the target gene, whereas the
steepness parameter $m$ is a measure of the collective effect of
groups of metabolite molecules and also defines the shape of the
Hill curve. This nonlinear parametrization describes the `biological
regulation process' that includes the production of the mRNA by
transcription of its structural gene, its possible intranuclear
processing by cleavage, its enzymatic degradation within the
nucleus, and its migration to the cytoplasm by some form of
diffusion or biological transport. Once in the cytoplasm, the mRNA
is both translated into the unstable enzyme and enzymatically
degraded.

System $\Gamma _{\rm biology}$ and its trivial chain generalization
in the linear part is considered to be a good model for the simplest
type of allosteric regulation in biochemistry, i.e., the inhibition
or activation of an enzyme or protein by a small regulatory molecule
that interacts with the enzyme at a site (allosteric site) other
than the active site at which catalytic activity occurs. The
interaction changes the shape of the enzyme, thus affecting the
active site of the standard catalysis. This change of shape of the
enzyme is sufficient to change its ability to catalyze a reaction in
either negative or positive way and enables a cell to regulate
needed metabolites. The allosteric regulation has the typical
features of a feedback loop in control theory if the regulatory
protein acts on the enzyme in the pathway of its own synthesis.



\section{The Nonlinear Observer} \label{non-obser section}

Many attempts have been made to develop nonlinear observer design
methods. One could mention the industrially popular extended Kalman
filter, whose design is based on a local linearization of the system
around a reference trajectory, restricting the validity of the
approach within a small region in the state space
\cite{Stephanopoulos},\cite{Wolovich}. The first systematic approach
for the development of a theory of nonlinear observers was proposed
some time ago by Krener and Isidori \cite{Krener 83}. In further
works, nonlinear transformations of the coordinates  have also been
employed to put the considered nonlinear sytem in a suitable
``observer canonical form'', in which the observer design problem
may be easily solved \cite{J.P. Gauthier 92},\cite{J.P. Gauthier
91},\cite{J.P. Gauthier 94}. The main idea in this case is to find a
state transformation to represent the system as a linear
differential equation plus a nonlinear term, which is a function of
the measured state.

\noindent In this section, we present the design of a nonlinear
software sensor in which the metabolite concentration is the
naturally measured state (the most easy to measure) and corresponds
to the mathematical state $X_3$ in the model introduced in the
previous section. Therefore, it seems logical to take $X_3$ as the
output of the system
\begin{eqnarray}
y=h(X)=X_3~.
\end{eqnarray}
We now apply the technique of high-gain observers that works for
many nonlinear systems and guarantees that the output feedback
controller recovers the performance of the state feedback controller
when the observer's gain is sufficiently high. The model given by
the aforementioned system $\Gamma _{\rm biology}$ has the form

\begin{eqnarray}
\Gamma _y: \ \ \ \left \{
\begin{array}{c}
\dot{X} = f(X)\\
 y = h(X)~,\\
\end{array}
\right. 
\end{eqnarray}

\noindent in which $X\in \mathbb{R}^3$, and moreover there is a
``physical subset'' $\Omega \subset \mathbb{R}^3$ where the system
lies. To make this mathematically precise we must introduce some
further mathematical terminology. Let us construct the $j$th time
derivative of the output. This can be expressed using Lie
differentiation of the function $h$ by the vector field $f$, $
{L_{f}}^{j}\left(h\right)\left(X(t)\right) $.
${L_{f}}^{j}\left(h\right)\left(X(t)\right)$ is the $j$th Lie
derivative of $h$ by $f$ and a function of $X$ defined inductively
as follows

\begin{eqnarray}
{L_{f}}^{0}\left(h\right)\left(X\right) &=& h\left(X\right)
\nonumber\\
\\
{L_{f}}^{j}\left(h\right)\left(X\right) &=& \frac{\partial}{\partial
X}\left( {L_{f}}^{j-1}\left(h\right)\left(X\right)\right)f(X)~.
\nonumber
\end{eqnarray}
When $\Gamma _y$ is observable, the map $\Phi : X \rightarrow
\Phi(X)$ is a diffeomorphism where

\begin{eqnarray}
 \xi =\Phi(X) = \left (
\begin{array}{c}
{L_{f}}^{0}\left(h\right)\left(X\right)\\
{L_{f}}^{1}\left(h\right)\left(X\right)\\
{L_{f}}^{2}\left(h\right)\left(X\right)
\end{array}
\right) = \left [
\begin{array}{c}
X_{{3}}\\
K_{{3}}X_{{2}}-{\it \gamma}_{{3}}X_{{3}}\\
K_{{3}} \left( K_{{2} }X_{{1}}-{\it \gamma}_{{2}}X_{{2}} \right)
-{\it \gamma}_{{3}} \left( K_ {{3}}X_{{2}}-{\it
\gamma}_{{3}}X_{{3}} \right)\\
\end{array}
\right ]~.
\end{eqnarray}

For $\Phi(X)$ to be a local diffeomorphism in a region $\Omega$, it
is necessary and sufficient that the Jacobian $\mathrm{d}\Phi(X)$
should be nonsingular on $\Omega$ and moreover that $\Phi(X)$ is
one-to-one from $\Omega$ to $\Phi(\Omega)$, see \cite{Shim}. Notice,
that no matter if we choose $H^{+}\left(X,\vartheta\right)$ or
$H^{-}\left(X,\vartheta\right)$, the coordinate transformation is
the same. This means that the structure of the observer will be the
same for both cases: gene activation or inhibition. \\

When the system is observable on $\Omega$, it can be rewritten in
the global coordinate system defined by $\Phi(X)$ in the following
matrix form:

\begin{eqnarray}
\Gamma'_{y}: \ \ \ \left \{
\begin{array}{c}
\dot{{\xi}} = F'\left({\xi}\right)= \left [
\begin{array}{c}
\dot{{\xi}}_1\\
\dot{{\xi}}_2\\
\dot{{\xi}}_3\\
\end{array}
\right ] = \left [
\begin{array}{c}
{{\xi}}_2\\
{{\xi}}_3\\
\varphi \left( {\xi} \right) \\
\end{array}
\right ] \\
y = C \ \xi = [1\ 0\ 0]\ \left [
\begin{array}{c}
{{\xi}}_1\\
{{\xi}}_2\\
{{\xi}}_3 \\
\end{array}
\right ]
\end{array} \right. ~,
\end{eqnarray}
where, moreover, $\varphi$ can be extended from $\Omega$ to the
entire $\mathbb{R}^3$ by a $\mathcal{C}^{\infty}$ function globally
Lipschitz on $\mathbb{R}^3$.
The latter form allows us to make use of the following result proven
by Gauthier and collaborators \cite{J.P. Gauthier 92}:\\

{\small \em Consider the system
\begin{eqnarray}
\Gamma _G: \, \dot{\hat{\xi}}=F'\left(\hat{\xi}\right)-S^{-1} C^T
\left( C \hat{\xi} - y \right)~, \label{observ coord trans}
\end{eqnarray}
where $S(\theta)$ is the solution of the matrix equation
\begin{equation}\label{Ricc}
\-\theta S-A^T S-S A+C^T C=0~,
\end{equation}
for $\theta$ large enough, with $A$ a matrix of Brunovsky form (
$A=\delta _{i,j+1}$; $\delta _{ij}$ is the Kronecker symbol), which
plays the role of a shift operator on $\mathbb{R}^n$.
Then, Eq.~ (\ref{observ coord trans}) defines an observer for
$\Gamma' _{y}$, with
\begin{eqnarray}
\left\| \hat{\xi}- \xi \right\| \leq M \exp \left(-\frac{\theta
}{3}\,t\right)\left\| \hat{\xi}_0- \xi_0 \right\|~.
\end{eqnarray}}
In our case, an observer is a dynamical system as given by
Eq.~(\ref{observ coord trans}) that Hill track the trajectory of the
original system (here $\Gamma' _{y}$). Notice that both systems are
identical unless an additional term that compensate the error in the
observer, where the error is given by the difference $\| \hat{\xi}-
\xi \|$, which is seen to be exponentially decreasing in time. The
Gauthier observer is particularly simple since it appears to be only
a copy of $\Gamma'_{y}$, together with a correction term that
depends only on the dimension of the state space and not on the
system $\Gamma'$ itself. In others words, the structure of the
observer does not
depend on the Hill steepness parameter $m$ (Eq.~(\ref{Eqn1})). 
\\

For the sake of concreteness we will construct the observer only for
the activation case. However, one should notice that only the
function $f\left(\hat{X}\right)$ will change for the inhibition
case.

The Gauthier observer in Eq.~(\ref{observ coord trans}) in the
original coordinates is given by
\begin{eqnarray}\label{Gobso}
\dot{\hat{X}}= f\left(\hat{X}\right)+\Upsilon
\left(\hat{X}\right)S^{-1}C^T\left( h(X) - h\left(\hat{X}\right)
\right)~,
\end{eqnarray}
where
\begin{eqnarray}
\left. \Upsilon \left(\hat{X}\right) =\frac{\partial
\Phi^{-1}}{\partial \hat{\xi}}\right|_{\hat{\xi}=\Phi(\hat{\xi})}~.
\end{eqnarray}
For the particular three-dimensional state space of $\Gamma _{\rm
biology}$ we get
\begin{eqnarray}\label{upspart}
\left. \Upsilon \left(\hat{X}\right)= \left[ \begin {array}{ccc}
{\frac {{\it \gamma}_{{2}}{\it
\gamma}_{{3}}}{K_{{2}}K_{{3}}}}&{\frac {{\it \gamma}_{{2}}+{\it
\gamma}_{{3}}}{K_{{2} }K_{{3}}}}&{\frac
{1}{K_{{2}}K_{{3}}}}\\\noalign{\medskip}{\frac {{ \it
\gamma}_{{3}}}{K_{{3}}}}&\frac{1}{K_{{3}}}&0\\\noalign{\medskip}1&0&0
\end {array} \right]~.\right.
\end{eqnarray}

The matrix $S(\theta)$ in the three dimensional case can be easily
computed by means of Eq.~(\ref{Ricc}) given in the Gauthier theorem
and
its inverse $S^{-1}(\theta)$ appears to be  
\begin{eqnarray}\label{invS}
S^{-1}(\theta)=
 \left[ \begin {array}{ccc} 3\,\theta&3\,{\theta}^{2}&{\theta}^{3}
\\\noalign{\medskip}3\,{\theta}^{2}&5\,{\theta}^{3}&2\,{\theta}^{4}
\\\noalign{\medskip}{\theta}^{3}&2\,{\theta}^{4}&{\theta}^{5}
\end {array} \right]~.
\end{eqnarray}

Plugging the matrices (\ref{upspart}) and (\ref{invS}) in
Eq.~(\ref{Gobso}), we get the following equation for the observer
introduced by Gauthier and collaborators as applied to our
biological case
\begin{eqnarray}\label{Gbiol}
\dot{\hat{X}}_{\rm biology}= f\left(\hat{X}\right)+ \left [
\begin{array}{c}
3\,{\frac {{\it \gamma}_{{2}}{\it
\gamma}_{{3}}\theta}{K_{{2}}K_{{3}}}} +3\,{\frac { \left( {\it
\gamma}_{{2}}+{\it \gamma}_{{3}} \right) {
\theta}^{2}}{K_{{2}}K_{{3}}}}+{\frac
{{\theta}^{3}}{K_{{2}}K_{{3}}}}\\
3 \,{\frac {{\it \gamma}_{{3}}\theta}{K_{{3}}}}+3\,{\frac
{{\theta}^{2}}{
K_{{3}}}}\\
 3\,\theta
\end{array}
\right ]\left( X_3 - \hat{X}_3 \right)~.\label{observ Originals
coord }
\end{eqnarray}

\begin{figure}[h]
  \hfill
  \begin{minipage}[t]{.45\textwidth}
    \begin{center}
      \includegraphics[height=4.5cm]{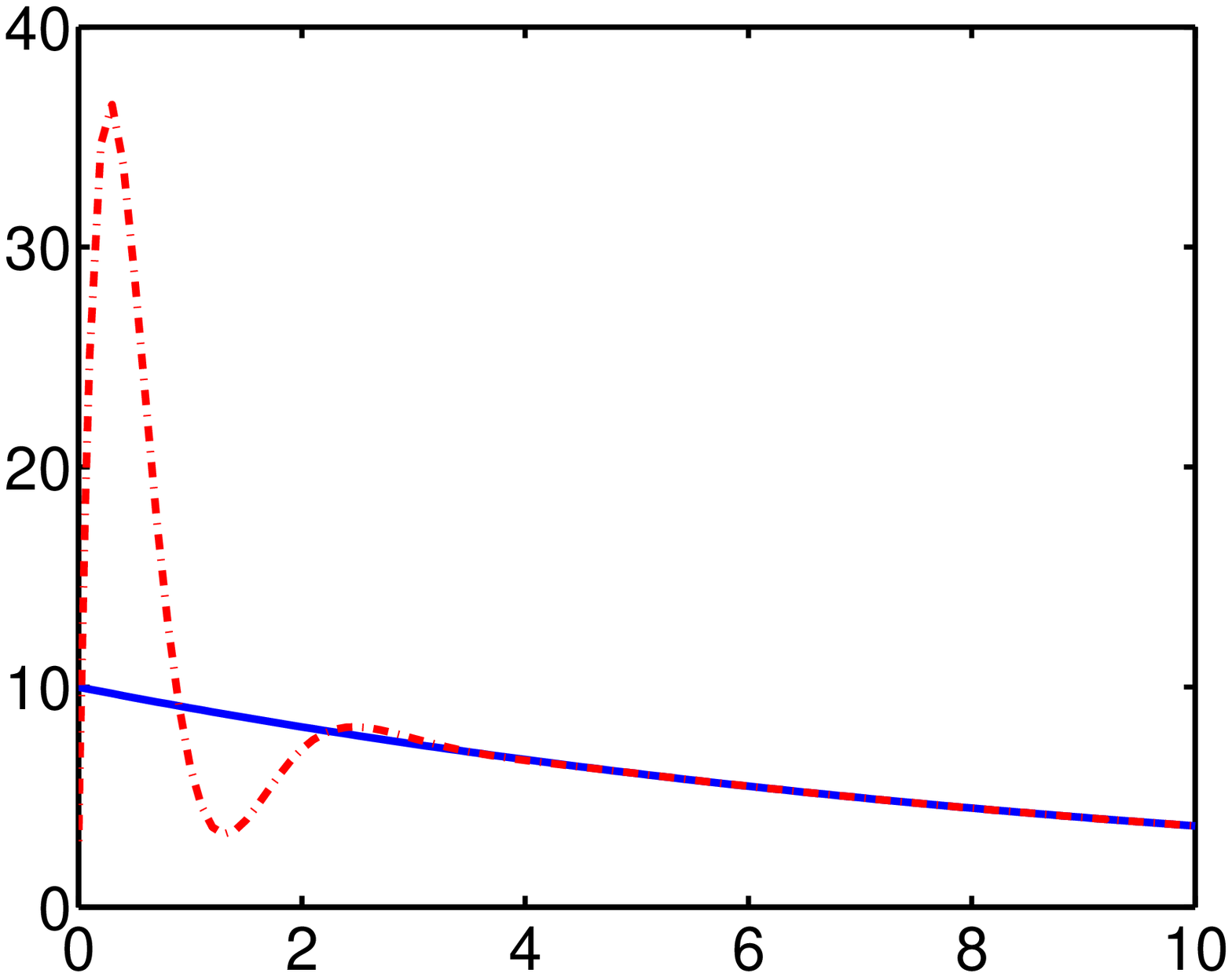}
      \put(-0,-10){time}
      \put(-140,110){$\hat{X_1}$}
      \put(35,110){$\hat{X_2}$}
       \put(-30,110){(a)}
      \put(-175,10){\rotatebox{90}{$X_1$ \small{(mRNA conc.)}}}
    \end{center}
  \end{minipage}
  \hfill
  \begin{minipage}[t]{.45\textwidth}
    \begin{center}
      \includegraphics[height=4.5cm]{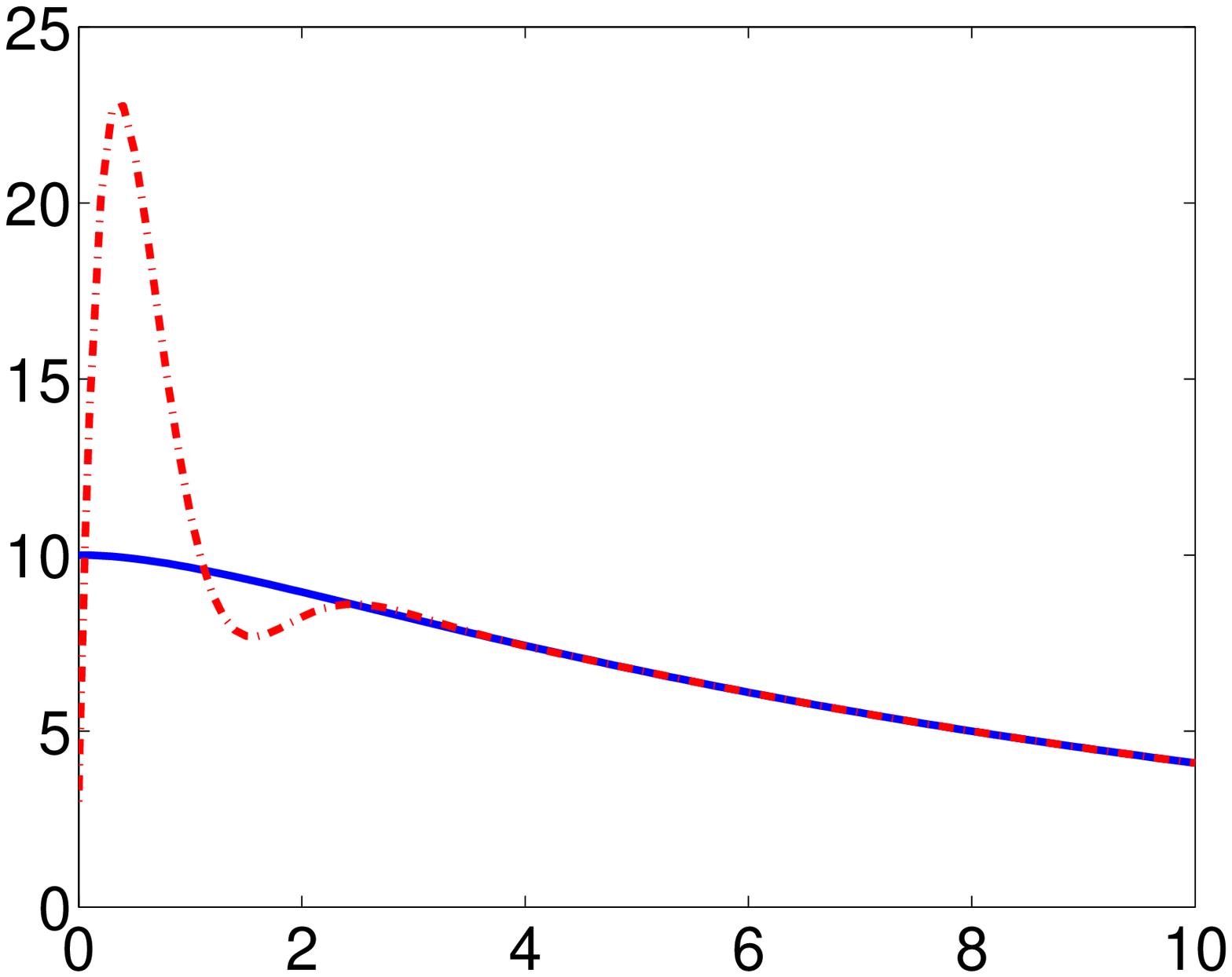}
       \put(-170,10){\rotatebox{90}{$X_2$ \small{(Protein conc.)}}}
       \put(-30,110){(b)}
    \end{center}
  \end{minipage}
  \hfill
  \caption{ The numerical simulation -- solid lines represent the true states and dotted
  lines represent the Gauthier estimates given by Eq.~(\ref{Gbiol}) for an activation case. Plot (a) represents the evolution of mRNA
   concentration in time and plot (b) the variation of
   protein (enzyme) concentration in time.}\label{simulaciones}
\end{figure}

We use this form of the Gauthier observer to estimate the states
$X_1$ and $X_2$ of the dynamical system $\Gamma _{\rm biology}$.
We work with
$\theta=1$ and the values of the parameters given in Table~1 that
are not necessarily the experimental values but are consistent with
the requirements of the model. Figure (\ref{simulaciones}) shows the
results of a numerical simulation, where the solid lines represent
the true states and the dotted lines stand for the estimates,
respectively. In addition, for the real system we have taken $m=2$
whereas for the observer $m=1$ in order to show the robustness of
this type of nonlinear observer with respect to the steepness
parameter.

\begin{table}[h]
\centering \caption{Variables and parameters} \label{tab:1}
\begin{tabular}{lll}
\hline\noalign{\smallskip}
Symbol & \ \ \ \ \ \ \ \ \ \ \ Meaning & Value (arb. units)  \\
\noalign{\smallskip}\hline\noalign{\smallskip}

$K_1$ & Production constant of mRNA &\ \qquad  $0.001$   \\
$K_2$ & Production constant of protein  &\ \qquad  $1.0$  \\
$K_3$ & Production constant of metabolite  &\ \qquad $1.0$  \\
$\gamma_1$ & Degradation constant of mRNA &\ \qquad $0.1$  \\
$\gamma_2$ & Degradation constant of protein &\ \qquad $1.0$  \\
$\gamma_3$ & Degradation constant of metabolite  &\ \qquad $1.0$  \\
$\vartheta$ & Hill's threshold parameter &\ \qquad $1.0$  \\
\noalign{\smallskip}\hline
\end{tabular}
\end{table}

\section{Conclusion}

We presented here the mathematical exercise of designing a high-gain
observer for a simple one-gene regulation dynamic process involving
end-product activation (inhibition leads to similar results), which
is able to rebuild in an effective way the non-measured
concentrations of mRNA and the involved protein. Thus, the
limitation of those experiments in which one has available only the
metabolite can be overcome by employing this simple observer. In
addition, this type of nonlinear observer could be used on line and
is robust with respect to $m$, i.e., it does not need the exact
value of the Hill steepness parameter. However, for more complex
inputs of more complicated observable dynamical systems, this
constant gain observer could have less performance and be overcome
by some {\em adaptive} observers that can change in order to work
better or provide more fit for a particular purpose. In the case of
more limited information, e.g., for unknown functional form of the
regulation function and high noise levels that can spoil the
performance of the observer, the completely different mathematical
procedure of creating dynamical extensions of the observer system
are required \cite{pre05}.

\section*{Acknowledgment}


\noindent This work was partially sponsored by grants from the
Mexican Agency {\em Consejo Nacional de Ciencia y Tecnolog\'{\i}a}
through project 46980.

\end{document}